\documentclass[%
aip,jap
amsmath,amssymb,
reprint,%
]{revtex4-1}

\usepackage{graphicx}
\usepackage{dcolumn}
\usepackage{bm}
\usepackage{amsmath}
\usepackage{amsfonts}
\usepackage{color} 

\begin{document}
\renewcommand{\i}{{\mathrm{i}}}
\newcommand{\e}{{\mathrm{e}}}
\newcommand{\kb}{\ensuremath{k_{ \text{\tiny B}}}}

\title[Andreev Current for low temperature thermometry]{Andreev Current for low temperature thermometry }

\author{T. Faivre}
\email{timothe.faivre@aalto.fi}
\affiliation{Low Temperature Laboratory, Departement of Applied Physics, Aalto University School of Science, POB 13500, FI-00076 AALTO, Finland
}%
\author{D. S. Golubev}%
\affiliation{Low Temperature Laboratory, Departement of Applied Physics, Aalto University School of Science, POB 13500, FI-00076 AALTO, Finland
}%
 \affiliation{Karlsruhe Institute of Technology (KIT), Institute of Nanotechnology, 76021 Karlsruhe, Germany}
\author{J. P. Pekola}
\affiliation{Low Temperature Laboratory, Departement of Applied Physics, Aalto University School of Science, POB 13500, FI-00076 AALTO, Finland
}%

\date{\today}

\begin{abstract}
We demonstrate experimentally that disorder enhanced Andreev current in a tunnel junction between a normal metal and a superconductor provides a method to measure electronic temperature, specifically at temperatures below 200 mK when aluminum is used. This Andreev thermometer has some advantages over conventional quasiparticle thermometers: for instance, it does not conduct heat and its reading does not saturate until at lower temperatures. Another merit is that the responsivity is constant over a wide temperature range. 
\end{abstract}

\maketitle
Thermometers are a cornerstone in experimental physics, from the premises of the thermodynamics to the latest observations of the universe involving bolometers\cite{Planck}. Yet in experiments in the millikelvin range and for objects of nanometer scales, it is hard to measure the actual temperature\cite{Pekola_JLTP}. Indeed, in mesoscopic transport measurement, local temperature variations can exist, self-heating needs to be avoided and the electronic temperature might deviate from the bath temperature. This leads to a need of special kinds of thermometers taking such constraints into consideration.  

Tunnel junctions have been used in this context for a long time \cite{review}, both with DC and more recently with fast readout\cite{Schmidt,Klaara}, but they suffer from various limitations. Operated at finite bias they heat (or cool) the circuit in which they are embedded in\cite{Juha}.  In addition, their responsivity tends to saturate at low temperature due to particular tunneling processes, such as environment assisted tunneling\cite{Pekola_dynes} or Andreev reflections\cite{Raj} . 

\begin{figure}[!htb]
\centering
\includegraphics[scale=0.33]{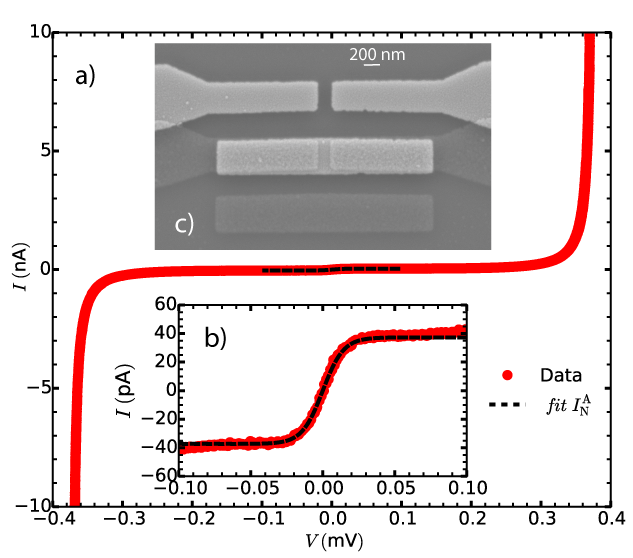}
\caption[SEM picture of the sample and its IV ]
{(a) {\it I-V} characteristics measured in 4-probe configuration. The sample is thermally anchored to the mixing chamber of a dilution refrigerator. The bath temperature $T_{\rm bath}=60$ mK is measured using a calibrated ruthenium oxide thermometer. The inset (b) shows a close up near zero bias, where the step due to the Andreev current is clearly visible. The dashed line is a fit using $I^A_{N}=37.5$ pA and $T_e\approx 80$ mK. The top inset (c) is a Scanning Electron Micrograph (SEM) of the sample.}
\label{fig:1}
\end{figure}

We are proposing to use the disorder enhanced Andreev current as a temperature probe. When a single electron attempts to leave the normal metal (N) trough a tunnel barrier (I), it is usually reflected if no states are available in the superconductor (S), i.e. if $eV < \Delta$.  Due to elastic scattering, this reflected electron may bounce around and finally make one or several attempts to tunnel. If the coherence time is long enough, all these attempts sum up coherently, and the rare case of being reflected as a hole (emitting a Cooper pair in S), becomes probable. This leads to a finite current, even for $eV < \Delta$. Measuring this current should give us a direct measure of the electronic temperature in the normal metal. 
The response is expected to be a linear function of temperature, which is an advantage, but the main reward is its operation point near zero bias. Therefore, such thermometer does not produce much self-heating, so one can easily think to use it in measurements where even tiny back-action from the thermometer to the system is  detrimental.
To demonstrate the principle of a thermometer based on Andreev current, we fabricated a sample device using electron beam lithography and shadow angle deposition. The junctions are formed by in-situ oxidation of the $H_S=20$ nm thick aluminum leads, before evaporating the island in the same chamber. The island is based on a titanium gold bilayer, so that the full island behaves as a normal metal. An aluminum/titanium buffer layers have been employed to ensure good contact between the different films. The island is finally Al(3nm)/Ti(2nm)/Au(10nm)/Ti(20nm). The resulting multilayer island is expected to play a role in enhancing the interference in the normal side of the junction, but a systematic study would be required to make definite statements. In the following we will treat the island as a single composite metal with a rectangular shape of length $L=2.73\: \mu {\rm m}$, width $W=0.4\: \mu {\rm m}$ and thickness $H=35\:  {\rm nm}$. The design maximizes the overlap area $\mathcal A$ of the junction, leading to $\mathcal{A}=0.5\: \mu \text{m}^2$ for a volume of $\mathcal{V}=0.038\: \mu \text{m}^3$. Figure \ref{fig:1}.c depicts the measured sample. This large area combined to the light oxidation (0.46 mbar of $\text{O}_2$ during 1 min 15 s) results in two series junctions, which we expect to have nearly equal normal state resistance $R_N=755 \: \Omega$.  Under this assumption the applied voltage is split equally between the junctions and $I_{\rm SINIS}(V)=I_{\rm NIS}(V/2)$ .

The current $I$ through one junction is expressed as  
\begin{equation}
\label{It}
I(V)=\frac{1}{2e}\int d\epsilon \  g(\epsilon) \left[ f (\epsilon-e V) -f (\epsilon+e V)   \right],
\end{equation}
 where we assume quasi-equilibrium on both sides of the junction, neglect charge imbalance on the superconducting side, 
and choose the distribution function in the normal island to be the equilibrium Fermi-Dirac one $f(E)=1/(e^{E/\kb T}+1)$. 
The energy dependent conductance $g(\epsilon)$ has the form (see, e.g. Ref. \onlinecite{CAR}): 
\begin{eqnarray}
\label{eq:gt}
g(\epsilon) &=&  g^{\rm BTK}(\epsilon)+ \frac{\theta(\Delta-|\epsilon|) \Delta^2}{\Delta^2-\epsilon^2} \frac{\Xi_N(2\epsilon) }{2e^2 \nu_N R_N^2 }
\nonumber\\ &&
+\, \frac{\Delta^2}{\Delta^2-\epsilon^2}\frac{\Xi_S \left[ 2W(\epsilon) \right] }{2e^2\nu_S R_N^2 },
\end{eqnarray}
where $\nu_{N},\nu_S$ are the densities of states in the normal and superconducting leads respectively, 
$W(\epsilon) = i \sqrt{\Delta^2-\epsilon^2}$ for $|\epsilon|<\Delta$, $W(\epsilon) = {\rm sgn}(\epsilon)\,\sqrt{\epsilon^2-\Delta^2}$ for $|\epsilon|>\Delta$
and  $g^{\rm BTK}(\epsilon)$ is the conductance of a junction connecting
two bulk leads derived by Blonder, Tinkham and Klapwijk (BTK)\cite{BTK} 
\begin{eqnarray}
\nonumber
g^{\rm BTK}(\epsilon)=\frac{e^2}{\pi\hbar}\sum_{n}  \left[ \frac{2T_n^2 \theta(\Delta-|\epsilon|)\Delta^2 }{T_n^2 \epsilon^2 +(2-T_n)^2 (\Delta^2-\epsilon^2)} \right. \\ 
\left. + \frac{2T_n \theta(|\epsilon|-\Delta) |\epsilon| }{T_n |\epsilon| +(2-T_n)\sqrt{\epsilon^2-\Delta^2}} \right] .
\end{eqnarray}
 $T_n$ are the set of transmission probabilities of the conducting channels. They relate to the junction normal state resistance $R_N$ by the Landauer formula $1/R_N=(e^2/\pi\hbar) \sum_n T_n$.

The disorder enhanced Andreev reflections are producing two corrections to the conductance $\propto\Xi_{N,S}(\omega)$ appearing in Eq. (\ref{eq:gt}). The functions $\Xi_{N,S}(\omega)$ are
expressed as double integrals of  the Cooperons $\mathcal{C}_{N,S}^{r,r'} (\omega)$ over the junction area ${\cal A}$,  
\begin{eqnarray}
\Xi_{N,S}(\omega) = \frac{1}{\mathcal{A}^2}  \int d^2r \int d^2r' {\rm Re} \left[ \mathcal{C}_{N,S}^{r,r'} (\omega)\right], 
\end{eqnarray}
and the Cooperons themselves are the solutions of the diffusion equation 
\begin{equation}
\label{eq:coo}
\left( -i \omega +1/\tau_{\varphi}^{N,S} - D_{N,S} \nabla^2 \right) \: \mathcal{C}_{N,S}^{r,r'}(\omega)=\delta(r-r').
\end{equation}
Here $\tau_{\varphi}^{N,S}$ are dephasing times in normal and superconducting leads respectively and
$D_{N,S}$ are the corresponding diffusion constants. 

Taking a typical value of dephasing time, $\tau_{\varphi}^N\sim 1$ ns, and estimating the diffusion constant, 
$D\simeq v_{\rm \tiny F} l/3\approx 2$ cm$^2$/s (here $v_{\rm \tiny F}\simeq 0.32\times 10^6$ m/s is Fermi velocity in titanium, and $l\sim 3$ nm as in Ref. \onlinecite{Ti}), 
we estimate the effective Thouless energy of the device to be
\begin{eqnarray}
\label{eq:ETh}
E_{\rm Th}\sim\frac{\hbar}{\tau_\varphi^N}+\frac{\hbar D}{L^2} \approx 1 \;\mu{\rm eV}.
\end{eqnarray}
This value is small as compared to typical bias voltages and temperatures, which
allows us to simplify the expression for the current (\ref{It}). Namely, since the function
$\Xi_{N}(2\epsilon)$ quickly decays for energies $|\epsilon|>E_{\rm Th}$, while
$\Xi_{S}(2W(\epsilon))$ varies slowly at energies $|\epsilon|<\Delta$, 
we can make the approximations (see \onlinecite{Appendix})
\begin{eqnarray}
\Xi_{N}(2\epsilon) &\approx & \frac{6\pi}{\mathcal{V}}\delta(\epsilon),
\nonumber\\
\Xi_{S}(2W(\epsilon)) &\approx & \frac{4}{\mathcal{A}H_S} \frac{1}{\sqrt{\Delta^2-\epsilon^2}},
\end{eqnarray}
where $H_S$ is the thickness of the superconducting film.
Substituting this result in Eq. (\ref{It}) for $eV\lesssim \Delta$, we arrive at the result\cite{Hekking,Hekking2}, 
\begin{eqnarray}
\label{eq:itot}
I(V)=I_{qp}(V)+I_N(V)+I_S(V),
\end{eqnarray}
where
\begin{eqnarray}
I_{\rm qp}(V)&=&\frac{1}{2e}\int d\epsilon  g^{\rm BTK}(\epsilon) \left[ f (\epsilon-e V) -f (\epsilon+e V)   \right],
\nonumber\\
I_N(V) &=& I_{N}^A \tanh\frac{eV}{2 \kb T}, \quad
I_S(V)= I_S^A \frac{eV}{\sqrt{\Delta^2-e^2V^2}}.
\end{eqnarray}  
The current amplitudes $I_{N,S}^A$ read
\begin{eqnarray}
\label{eq:IA}
I_N^A = \frac{3\pi}{2}\frac{\hbar}{e^3\nu_N{\cal V}R_N^2},\;\; I_S^A = \frac{2\hbar}{e^3\nu_S{\cal A} H_SR_N^2}.
\end{eqnarray}
Assuming $\nu_N\approx7. \: 10^{47} \: {\rm J}^{-1}\: {\rm m}^{-3}$ as the titanium density of states, one gets $I_S^A = 19$ pA and $I_N^A = 7.25$ pA with the parameters of our sample device. The experimental value of $I^A_{N}=$ 37.5 pA is obtained by fitting the low bias region of the {\it I-V} characteristics at base temperature (dashed black line of Fig. \ref{fig:1}). The discrepancy between the experimental value and its theoretical prediction might be a consequence of the layered island as the two values would coincide if one considers an effective thickness of the normal island to be $H=7.4$ nm instead of $35$ nm. 

At low temperature ($\kb T \ll \Delta$) and for weakly transparent junctions ($T_n\ll 1$), one can approximately express zero  bias conductance in the form
\begin{eqnarray}
\label{eq:G}
G_0(T)= \frac{1}{R_N}  \left(  \gamma +\sqrt{\frac{2\pi\Delta}{\kb T}} \e^{-\Delta/\kb T} \right) + \frac{e I_{N}^A}{2\kb T},
\end{eqnarray}
where 
\begin{eqnarray}
\gamma = R_N \frac{e^2}{\pi\hbar}\sum_{n} \frac{2T_n^2}{(2-T_n)^2} +  \frac{e I_{S}^AR_N}{\Delta}
\end{eqnarray}
is an effective Dynes parameter accounting for the sub-gap leakage current. 
Environment assisted tunneling\cite{Pekola_dynes} may also contribute to the phenomenological parameter $\gamma$. Nevertheless, with the value of $I_S^A=19$ pA estimated above, one gets $e I_{S}^AR_N /\Delta \approx 0.65 \times 10^{-4}$ which agrees well to $\gamma$ $= 0.8 \times 10^{-4}  $ in the experiment. 

\begin{figure}[h]
\centering
\includegraphics[scale=0.33]{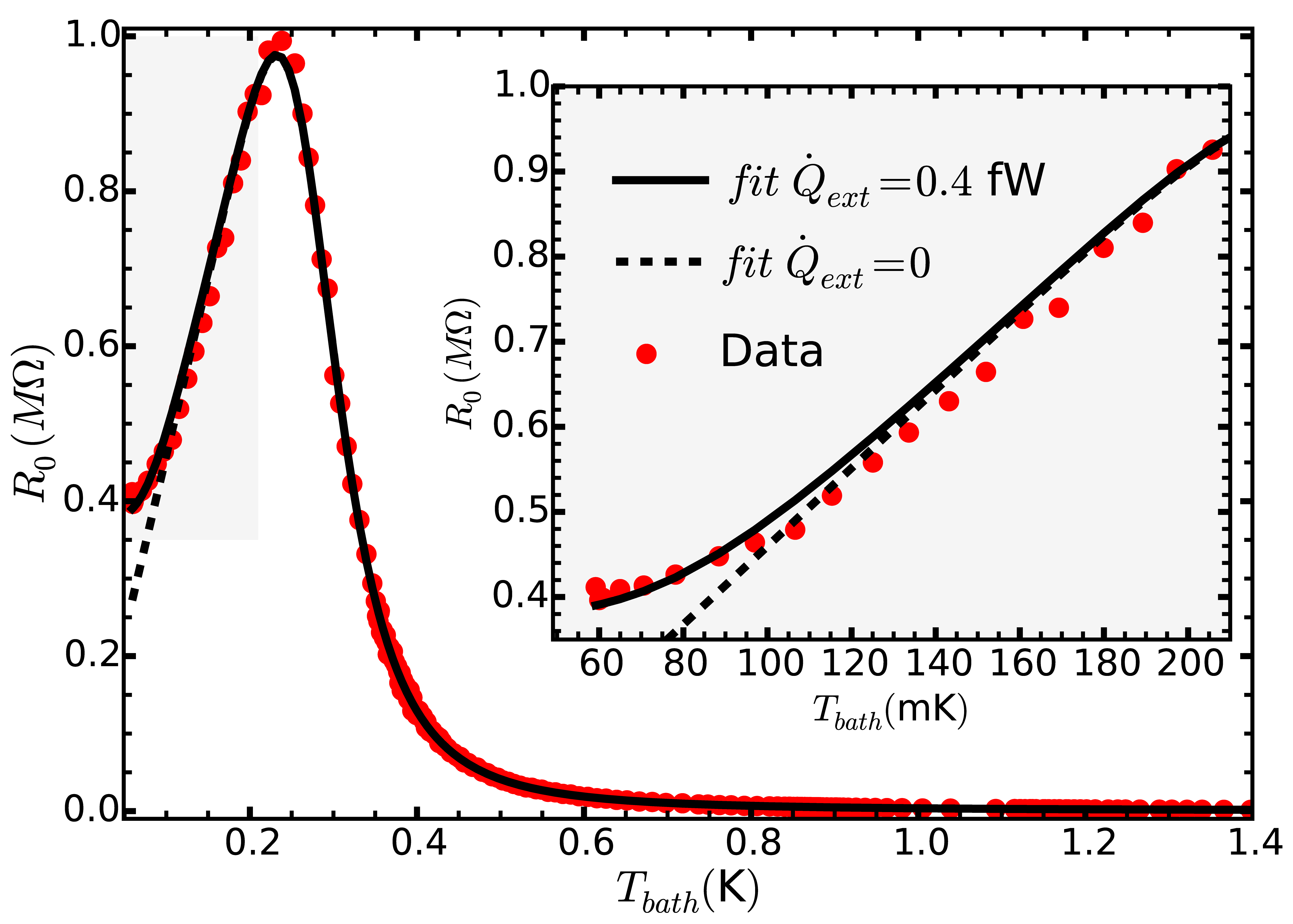}
\caption[zero bias resistance]
{Zero bias resistance as a function of temperature. The fitting model, indicated by black lines, uses Eqs. (\ref{eq:G}) and (\ref{eq:hbe}). The parameters are $R_N=755 \ \Omega$, $\Delta=221.7 \: \mu eV$, $\gamma=0.8\times10^{-4}$, $I^A_{N}=37.5$ pA. One more parameter is the external heat current $\dot{Q}_{\rm ext}$ accounting for the deviation observed at low temperature (see inset).}
\label{fig:2}
\end{figure}

Figure  \ref{fig:2} depicts the zero bias resistance, extracted from the {\it I-V} characteristics by numerical differentiation. The aluminum gap $\Delta=221.7 \: \mu e\text{V}$ is used as a fitting parameter at temperatures near the aluminum transition, where the total conductance is dominated by the quasiparticle tunneling. The responsivity in this regime where $k_B T\ll \Delta$ but where quasiparticle conductance still dominates reads  
\begin{eqnarray}
\label{eq:respqp}
\frac{\partial R_0}{\partial T}=-\frac{R_N}{T}\sqrt{\frac{\Delta}{2\pi \kb T}} \e^{\Delta/\kb T} \:.
\end{eqnarray}
The NIS thermometer responsivity will eventually vanish due to the sub-gap leakage, as the NIS resistance reaches the limit set by the $\gamma$ parameter.

The responsivity changes sign for the temperature $\kb T_{0}\sim \Delta/\ln{\left( \frac{\Delta}{eI^A_NR_N}\right) }$ . For temperatures below $T_0$ Andreev current is dominating the zero bias conductance, leading to the positive responsivity 
\begin{eqnarray}
\label{eq:respa}
\frac{\partial R_0}{\partial T}=\frac{2\kb }{eI_N^A}
\end{eqnarray}  
which remains constant for $\kb T>E_{\rm Th}$.
This phenomenon is also called reentrance effect, it has been predicted in the context of NIS junction theoretically\cite{Tanaka,Nazarov} and measured experimentally\cite{Quirion,Reulet}. 

In order to reproduce the data presented in Fig. \ref{fig:2} theoretically, we need to take into account that the electronic temperature saturates at around 85 mK when cooling the bath ($T_{\rm bath}$) below this temperature. This can be incorporated in the modeling by considering the normal island of the device to be a free electron gas in quasi-equilibrium so that its temperature $T_e$ is the solution of the following Heat Balance Equation (HBE):
\begin{equation}
 \dot{Q}_{\rm a} + 2\ \dot{Q}_{\rm qp}(V,T_e)+\dot{Q}_{\rm ext}=\Sigma \mathcal{V} (T_e^\alpha-T_{\rm bath}^\alpha) \: .
\label{eq:hbe}
\end{equation}
Although Andreev current does not transport heat across the barrier, it has been shown that it produces power $\dot{Q}_{\rm a}= I_N \times V$ into the normal island\cite{Averin,Vasenko}. $\dot{Q}_{\rm qp}(V,T_e)$ represents the heating (or cooling) due to quasi-particle transport through one of the junctions. One can neglect this term near zero bias, but it dominates the left-hand side of Eq. (\ref{eq:hbe}) as soon as $eV$ approaches $\Delta$. The term on the right-hand side is the electron-phonon coupling, with $\alpha=5$ and $\Sigma \sim 1\: {\rm nW}/{\rm K}^5/\mu{\rm m}^3$ for most metals\cite{review}. The phonon temperature is assumed to follow the bath temperature $T_{\rm bath}$ so that assuming an external heat load  $\dot{Q}_{\rm ext} = 170\: \text{aW} $, the electronic temperature saturates around $T_{\rm sat}=\sqrt[5]{\dot{Q}_{\rm ext}/\Sigma \mathcal{V}}= 85$ mK.

Although not less than 5 parameters are needed to describe the zero bias resistance as a function of temperature fully, only a single one ($I^A_N$) accounts for the constant responsivity in the range from $ E_{\rm Th}/ \kb $ to $T_0 $. Furthermore, this parameter can be known beforehand, by measuring a single {\it I-V} characteristic.

\begin{figure}[h]
\centering
\includegraphics[width=1.\linewidth, height=.62\linewidth]{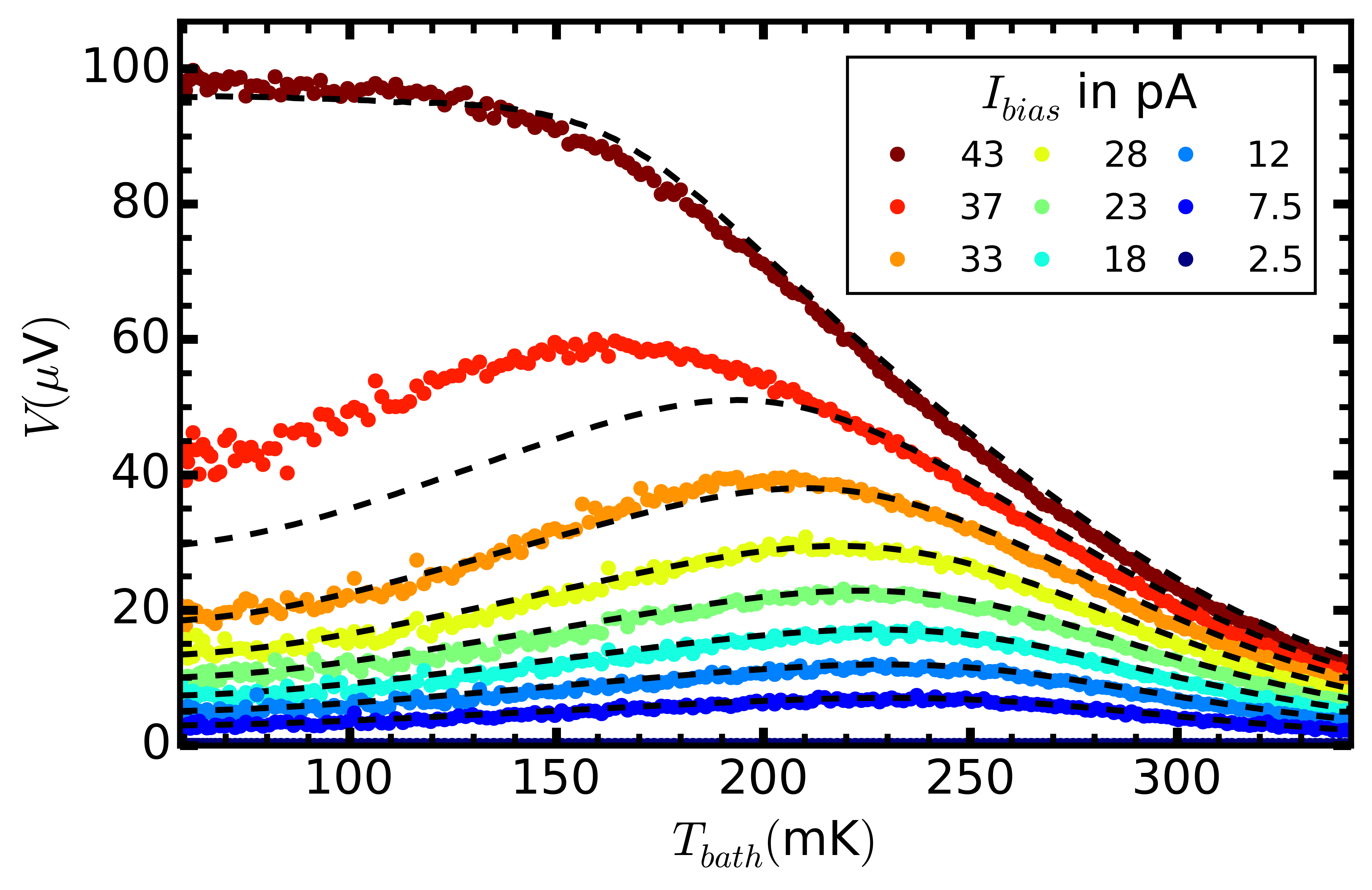}
\caption[voltage response]
{Voltage response of the thermometer. The dashed lines are calculated using the parameters listed in Fig. \ref{fig:2} caption. The non-monotonic behavior of the voltage denotes two different regimes, quasiparticle NIS tunneling at high temperature and disorder enhanced Andreev current at low temperature.}
\label{fig:3}
\end{figure}

Let us now extract the responsivity and sensitivity of our implementation of an Andreev thermometer.  
Figure \ref{fig:3} shows the voltage response of a current biased SINIS device. Parameters which have been extracted previously are re-used to plot the theoretical response (dashed lines) using Eqs. (\ref{It}) and (\ref{eq:hbe}), excluding self heating. The upper curve ($I_{\rm bias}=43 \: {\rm pA} > I^A_{N}$) is a typical response of a NIS quasiparticle thermometer. The saturation, due to the $\gamma$ parameter, is limiting the range of this measurement, as suggested by Eq. (\ref{eq:gt}). On the contrary we do not expect saturation of the Andreev thermometer, whose maximum responsivity is given by   $\partial V / \partial T_{\rm max}=2\kb/e \simeq 172\: \mu{\rm V/K} $. 
Due to the positive responsivity, implementing electro-thermal feedback requires a voltage biased device, but the relatively small logarithmic
derivative of the resistance, $\alpha=d \ln R/d\ln T \approx 1$ will limit its strength. 
In order to reduce self heating one can operate the thermometer close to zero bias voltage  and measure the impedance of the structure. The responsivity is then $\left. \frac{\partial R}{\partial T}\right|_{0} = 4.28 \: \text{M}\Omega /\text{K}$, which is in good agreement with the theoretical prediction given by Eq. (\ref{eq:respa}).

In the small bias regime, the main source of noise comes from the voltage pre-amplifier. The Noise Equivalent Temperature (NET) of the thermometer can then be directly calculated as 
\begin{equation}
\text{NET} \equiv \frac{S_v}{\partial V/\partial T} \  = 20.8\: \mu \text{K}/\sqrt{\text{Hz}}
\end{equation}
where $S_v$ is the input noise of the voltage amplifier, typically $S_v \approx 1 {\rm nV}/\sqrt{{\rm Hz}}$.
If one can assume the noise of the amplifier to depend only weakly on the sample impedance, the NET is expected to be constant over the full temperature range. 

The Noise Equivalent Power (NEP) is related to the NET as
$\text{NEP}= G_{{\rm th},\Sigma} \times NET$, where $G_{{\rm th},\Sigma} $ is the total thermal conductance, linearized near the working point. 
Near zero bias voltage the electron-phonon coupling dominates the heat transport and $G_{{\rm th},\Sigma}\approx 5 \Sigma V T^{4}$. We then expect $\text{NEP}=1 \times 10^{-18} \ \text{W}/\sqrt{\text{Hz}}$ at 100 mK, with $\Sigma =2.4\: {\rm nW}/{\rm K}^5/\mu{\rm m}^3$, which is the value of bulk gold. We believe this estimation of the NEP to be the worst case estimate as the island is composed partly of titanium with a smaller $\Sigma$ value. As the responsivity of the Andreev thermometer is constant, reducing the temperature leads to a quick improvement of the NEP. At 20 mK we would expect $\text{NEP}=1.6 \times 10^{-21} \ \text{W}/\sqrt{\text{Hz}}$ within this model.


Defining a figure of merit for a thermometer is a problem with an application-specific answer\cite{Pekola_JLTP}.
According to Eq. (\ref{eq:respa}) the main advantage of Andreev thermometer appears to be its constant responsivity over a wide range of temperature, bounded below by the Thouless energy (Eq. (\ref{eq:ETh})). A realistic estimation gives $2.5\sim 25$ mK depending on the value of  $\tau_{\varphi}^{N} = 1.5\sim 0.15$ ns measured in other experiments\cite{pothier,Pierre}.  Once the Andreev current $I^A_N$ has been measured at a single bath temperature, the responsivity is simply a number involving only fundamental constants. We measured the responsivity in our device to be $\left. \partial R/\partial T \right|_{0} = 4.28 \: \text{M}\Omega /\text{K}$, which was constant over the full temperature range of the measurement [80--200 mK]. 
Generally, in an equilibrium environment where $\dot{Q}_{\rm ext}$ can be neglected, the saturation of the electronic temperature is determined by self heating. In this respect Andreev thermometer is favorable since it operates near zero bias.  This is not the case for instance for a NIS thermometer for which a trade-off between the responsivity and the operation range cannot be avoided. 

As Andreev thermometer is probing the local temperature of a metallic island, one can expect this system to be a radiation absorber, and we estimate the NEP to be $1 \times 10^{-18} \ \text{W}/\sqrt{\text{Hz}}$ at 100mK. 
Further measurements at lower temperature, by reducing the external heat load of the present experiment would be required to test the low temperature limitations of an Andreev thermometer.

This material is based upon work supported by the
Academy of Finland under projects no. 139172 and
250280 (LTQ Centre of Excellence), and by the European
Commission under project no. 264034 (Q-NET Marie
Curie Initial Training Network). The
research made use of OtaNano, the Otaniemi Research Infrastructure
for Micro- and Nanotechnology.

\end{document}